\documentclass[amsmath,amssymb,pra,final,showpacs,twocolumn]{revtex4-1}

\usepackage{bm,hyphenat,xspace}
\usepackage{graphicx,epsfig}

\newcommand {\mcu}{\mathcal{U}}

\begin{document}

\title {Universal bosonic tetramers of dimer-atom-atom structure}

\author{A.~Deltuva} 
\affiliation{Centro de F\'{\i}sica Nuclear da Universidade de Lisboa, 
P-1649-003 Lisboa, Portugal }

\received{\today}

\begin{abstract}
Unstable four-boson states having an approximate dimer-atom-atom structure
are studied  using momentum-space 
integral equations for the four-particle transition operators. 
For  a given Efimov trimer the universal
properties of the lowest associated tetramer are determined.
The impact of this tetramer on the atom-trimer and dimer-dimer
collisions is analyzed. The reliability of the three-body
dimer-atom-atom model is studied.
\end{abstract}

 \maketitle

\section{Introduction}

It took more than 30 years till 
Efimov's prediction for the existence of weakly
bound three-particle states  with asymptotic discrete scaling symmetry
 \cite{efimov:plb} was confirmed in the cold-atom physics
experiments \cite{kraemer:06a}. This discovery raised the interest in 
few-body systems with resonant short-range interactions,
both experimentally and theoretically.
After the three-body system where semi-analytical
results have been obtained as summarized in Ref.~\cite{braaten:rev},
the next step in the complexity
to explore the Efimov physics is the system of four identical bosons
where several numerical studies 
\cite{lazauskas:he,hammer:07a,stecher:09a,schmidt:10a,yamashita:10a,deltuva:10c,stecher:11a}
are already available.
Although there is no four-boson Efimov effect \cite{amado:73a,wang:12a},
 the properties of the four-boson system are strongly
affected by the three-boson Efimov effect.
The observables of the four-boson collisions 
 show the same  discrete scaling symmetry
and are related to the respective trimer binding energies in a 
universal way  provided the size of the involved few-body bound states 
greatly exceeds the interaction range \cite{stecher:09a,deltuva:10c}.
Furthermore,  below each Efimov trimer in a certain regime
there are two tetramers \cite{hammer:07a,stecher:09a}.
The tetramers associated with the excited trimers are
unstable bound states (UBS) \cite{res_cpl,deltuva:11a,stecher:11a} 
and lead to resonant effects in the four-boson reactions.
First numerical calculations of the 
four-boson recombination  \cite{stecher:09a}
and dimer-dimer collisions  \cite{dincao:09a}
in the adiabatic hyperspherical representation
were followed by the work of Refs.~\cite{deltuva:10c,deltuva:11b,deltuva:12a}
employing the  integral equations for the transition operators
that were solved in the momentum-space framework.
In the latter case the  universal limits for the scattering observables
and tetramer properties such as widths and intersections with thresholds 
were obtained with considerably higher accuracy;
furthermore, remarkable resonant effects were predicted in 
the atom-trimer collisions as well \cite{deltuva:11a}.
The theoretical results \cite{stecher:09a,dincao:09a,deltuva:11b,deltuva:12a}
are  roughly consistent with
the existing data from  the dimer-dimer and four-atom recombination 
experiments \cite{ferlaino:08a,ferlaino:09a,pollack:09a,zaccanti:09a}
performed in a regime that is not strictly universal.

\begin{figure}[!]
\includegraphics[scale=0.5]{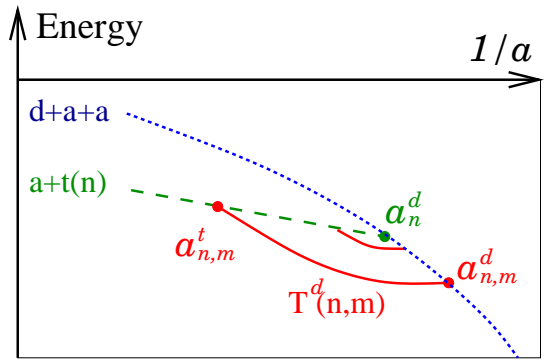} 
\caption{\label{fig:ae} (Color online)
Schematic representation of the four-boson energy spectrum
in the vicinity of the  dimer-atom-atom (dotted line)
and the $n$th atom-trimer (dashed line) threshold intersection
at $a= a_n^d$. The tetramers (solid lines)
accumulate at  $a= a_n^d$ and $E=-b_d$ but
for simplicity only two states ($m=0,1$) are displayed;
only qualitative relations are preserved. The  $(n,m)$th tetramer 
intersects the dimer-atom-atom and the  $n$th atom-trimer thresholds
at $a= a_{n,m}^d$ and $a= a_{n,m}^t$, respectively.
}
\end{figure}

In the present work we will study a different type of bosonic tetramers,
namely, the ones of the approximate dimer-atom-atom structure. 
Their existence was predicted in Refs.~\cite{braaten:rev,dincao:09a} as a
consequence of the three-body Efimov effect in the three-body system made off 
a dimer and two atoms.
This is illustrated in Fig.~\ref{fig:ae} where we schematically show
the four-boson energy ($E$) spectrum as function of $1/a$ with $a$ being the
two-boson scattering length.
We consider the regime around the special value $a = a_n^d$ that corresponds
to the $n$th Efimov trimer being at the atom-dimer threshold, i.e.,
$b_d = b_n$ where $b_d$ ($ b_n$) is the dimer ($n$th trimer)
binding energy relative to the free particle threshold.
If the trimer is a true bound state with zero width, i.e.,
if deeper dimers are absent (which is not the case in typical
experiments \cite{ferlaino:08a,ferlaino:09a,pollack:09a,zaccanti:09a}),
the atom-dimer  scattering length $A_d$ is infinite at $a = a_n^d$.
 Sufficiently close to $E=-b_n$ and $a = a_n^d$
where $A_d$ greatly exceeds the dimer size being
of the order of $a$, one may expect 
to mimic some properties of the few-boson systems  
using a model that considers the dimer as a pointlike particle.
Then in the effective three-body system consisting of a dimer and two atoms
there are two atom-dimer pairs with infinite two-body scattering length $A_d$.  
In such a three-body model of the four-boson system
the three-body Efimov effect occurs, however,  with a very large discrete 
scaling factor $e^{\pi/s_0} \approx 2.016 \times 10^5$ \cite{braaten:rev}.
The resulting Efimov states accumulate at  $E = -b_d$.
Their dimer-atom-atom structure is only an
approximation but no attempt has been made so far to describe them
rigorously as four-boson states. This will be done in the
present work using exact four-particle equations. 
We label the tetramers of this type  
by two integers $(n,m)$ where $n$ refers to the
associated Efimov trimer and $m=0,1,2,\ldots$ distinguishes between
different states existing around  $a = a_n^d$ with  fixed $n$.
The $(n,m)$th tetramer intersects the  $n$th atom-trimer threshold
at $a= a_{n,m}^t$ and the dimer-atom-atom threshold at $a= a_{n,m}^d$.
The calculation of these tetramers is technically very difficult task
owing their weak binding and very large $e^{\pi/s_0}$; our results  
will be limited to $m=0$ states.
Furthermore, all these tetramers  lie above the dimer-dimer threshold
and above all lower atom-trimer thresholds $n'$ with 
$n' < n$. Therefore the considered tetramers are UBS  with finite width.
We will extract their properties using rigorous four-particle scattering
calculations. Performing in addition the three-body calculations
we will establish the limitations of the  dimer-atom-atom model.

In Sec.~\ref{sec:th} we shortly recall the technical framework.
In Sec.~\ref{sec:res} we present results for tetramer
properties and their effect on the atom-trimer and dimer-dimer
 scattering observables.
 We summarize in  Sec.~\ref{sec:sum}.

\section{Four-boson scattering equations \label{sec:th}}

To describe the collisions in the four-boson system we
use exact Alt, Grassberger and Sandhas (AGS) 
equations \cite{grassberger:67} for the transition operators 
$\mcu_{\beta\alpha}$ in the symmetrized form 
\begin{subequations} \label{eq:U}
\begin{align}  
\mcu_{11}  = {}&  P_{34} (G_0  t  G_0)^{-1}  
 + P_{34}  U_1 G_0  t G_0  \mcu_{11} + U_2 G_0  t G_0  \mcu_{21} , 
\label{eq:U11} \\  
\mcu_{21}  = {}&  (1 + P_{34}) (G_0  t  G_0)^{-1}  
+ (1 + P_{34}) U_1 G_0  t  G_0  \mcu_{11} , \label{eq:U21} \\
\mcu_{12}  = {}&  (G_0  t  G_0)^{-1}  
 + P_{34}  U_1 G_0  t G_0  \mcu_{12} + U_2 G_0  t G_0  \mcu_{22} , 
\label{eq:U12} \\  
\mcu_{22}  = {}&    
(1 + P_{34}) U_1 G_0  t  G_0  \mcu_{12} . \label{eq:U22}
\end{align}
\end{subequations}
Here $G_0$ is the free resolvent,  $P_{34}$ is the  
permutation operator of particles 3 and 4,
$t = v+vG_0t$ is the two-particle transition matrix
derived from the potential $v$, 
and $U_{\alpha}$ are the symmetrized AGS operators \cite{deltuva:07a}
for  $3+1$ ($\alpha=1$) and $2+2$ ($\alpha=2$) subsystems.
We solve the AGS equations \eqref{eq:U}
in the momentum-space framework using the 
partial-wave decomposition. In the present study only the  
states with zero total angular momentum and positive parity ($0^+$)
corresponding to the quantum numbers of the tetramers need to be considered;
among them  the states with the
nonconserved angular momentum of the three-boson subsystem
$J \le 3$ have to be included for the convergence. 
The atom-atom interaction $v$ is taken over from Ref.~\cite{deltuva:10c};
it is given by a separable $S$-wave potential with
one- or two-term gaussian form factors.
Further technical details can be found in 
Refs.~\cite{deltuva:12a,deltuva:07a,deltuva:ef}.

The scattering amplitudes for all elastic and inelastic
two-cluster reactions are obtained as the on-shell matrix elements of 
the AGS transition operators  \cite{deltuva:07a};
the amplitudes for breakup and recombination \cite{deltuva:12a}
are given by the integrals involving $\mcu_{\beta\alpha}$.

The unstable tetramers manifest themselves as poles of the 
operators $\mcu_{\beta\alpha}$ in the complex energy plane
at $-B_{n,m}^d - i\Gamma_{n,m}^d/2$ with
 $-B_{n,m}^d $ being the energy of the $(n,m)$th tetramer
relative to the four-boson breakup threshold and 
$\Gamma_{n,m}^d$ being its width.
The UBS pole in the complex energy plane
is located in one of the unphysical sheets that 
is adjacent to the physical sheet \cite{res_cpl} and
affects the physical observables 
leading to resonant effects in the four-boson scattering. 
Thus, the properties of the
 unstable tetramers can be extracted from the behavior of
$\mcu_{\beta\alpha}$ as described in Ref.~\cite{deltuva:11a}.

In the three-body dimer-atom-atom (3BDAA) model
the corresponding reactions are
described by solving the AGS three-body scattering equations \cite{alt:67a}.
These calculations need two pair potentials.
The atom-atom potential $v_{aa} = v$ is the same as in full four-boson
calculations. The atom-dimer potential  $v_{ad}$ is taken in the momentum-space
representation as $\langle \mathbf{k}' | v_{ad} | \mathbf{k} \rangle 
= \exp(-{k'}^2/\Lambda_d^2) \, \lambda_d \exp(-{k}^2/\Lambda_d^2)$.
Its strength $\lambda_d$ and momentum cutoff (inverse range)
parameter $\Lambda_d$ for each $a$
are adjusted to reproduce $A_d$, $(b_n-b_d)$ and atom-dimer effective range 
predicted by the original three-boson calculations using $v_{aa}$.
 As a consequence,  $v_{ad}$ yields an accurate description
of the low-energy atom-dimer scattering.
We note that  the range of $v_{ad}$ with
$1/\Lambda_d \approx a$ greatly exceeds the range of $v_{aa}$
that is much smaller than $a$.

\section{Results \label{sec:res}}

Unless stated otherwise, our results refer to rigorous four-boson
calculations.
We present them as dimensionless ratios that are
independent of the short-range interaction details in the universal limit.
For this one needs
to consider reactions involving highly excited Efimov trimers 
where the finite-range effects become negligible.
As shown in previous calculations
\cite{schmidt:10a,deltuva:10c,deltuva:11a,deltuva:11b,deltuva:12a}, 
$n\ge 3$ is sufficient for high accuracy (note that our nomenclature
starts with $n=0$ for the ground state).
This is fully consistent with our present results. For example,
for the $(n,0)$th tetramer intersection with the corresponding
atom-trimer threshold we obtain 
$a_{3,0}^t/a_3^d \approx 1.609$ and $a_{4,0}^t/a_4^d \approx 1.608$
using one-term form factor in the potential $v$ and 
$a_{3,0}^t/a_3^d \approx 1.607$ and $a_{4,0}^t/a_4^d \approx 1.608$
using two-term form factor, respectively. 
Thus, we conclude that for large $n$ the universal ratio is
\begin{equation} \label{eq:at}
a_{n,0}^t/a_n^d = 1.608(1),
\end{equation}
corresponding to $A_d/a_{n,0}^t = 5.491(3)$.
The calculation of  the  tetramer intersection with the
dimer-atom-atom threshold, $a_{n,0}^d$, 
 turns out to be numerically more complicated and 
less accurate due to its proximity to $a_n^d$. We have 
found that $a_{n,0}^d/a_n^d > 0.9999$.
The  higher tetramers $(n,m \ge 1)$ may exist only at extremely large 
 atom-dimer  scattering length $A_d$, i.e., extremely close to $a=a_n^d$,
and cannot be resolved in our four-boson calculations. Thus, with increasing $m$
the ratios $a_{n,m\ge 1}^t/a_n^d$ $(a_{n,m}^d/a_n^d)$ very rapidly
approach 1 from above (below). For example, 
our estimations using the 3BDAA model are
\begin{equation} \label{eq:ad}
1- a_{n,0}^d/a_n^d = 1.2(1) \times 10^{-6},
\end{equation}
corresponding to $A_d/a_{n,0}^d = -1.8(1)\times 10^6$, and
\begin{equation} \label{eq:at1}
a_{n,1}^t/a_n^d -1 = 1.3(1) \times 10^{-6},
\end{equation}
corresponding to $A_d/a_{n,1}^t = 1.7(1)\times 10^6$ and
$(b_n-b_d)/b_d = 2.6(2) \times 10^{-13}$.

\begin{figure}[!]
\includegraphics[scale=0.64]{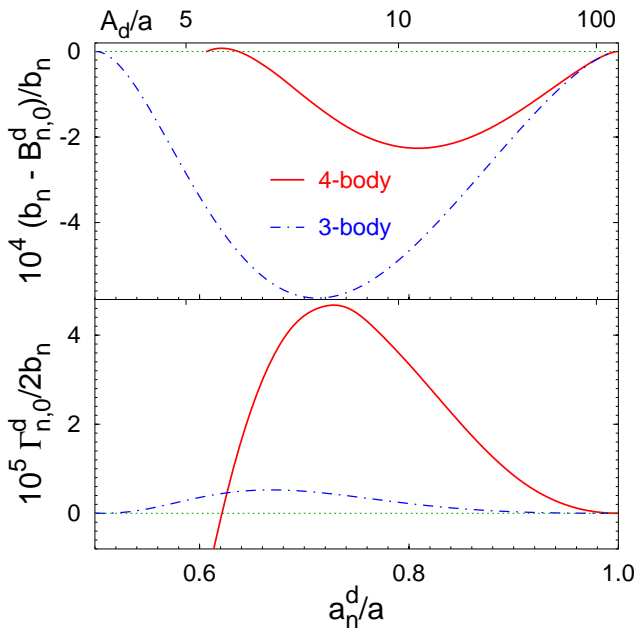} 
\caption{ (Color online)
Position of the $(n,0)$th tetramer relative to the 
$n$th atom-trimer threshold  (top) and its width (bottom)
as functions of the two-boson scattering length $a$ and
 the atom-dimer scattering length $A_d$.
The results of the full four-body (solid curves) and the 3BDAA
(dashed-dotted curves) models are compared.
The thin dotted line represents the zero value.}
\label{fig:BG}
\end{figure}

The universal values \eqref{eq:at} and \eqref{eq:ad} indicate
that the $(n,0)$th tetramer as UBS exists at 
$0.6219 < a_n^d/a < 1.0000012$.
We show in Fig.~\ref{fig:BG} the  $a$-evolution of the $(n,0)$th 
tetramer properties, i.e., its relative distance to the 
$n$th atom-trimer threshold $(b_n-B_{n,0}^d)/b_n$ and its width 
$\Gamma_{n,0}^d/2b_n$ as dimensionless universal quantities.
We include also the predictions of the  3BDAA model
and demonstrate that it is reliable only for the tetramer
position at $  a_n^d/a > 0.96$ where $|A_d|/a > 50$.
Although the 3BDAA model describes well the $n$th trimer binding and
the low-energy atom-dimer scattering, it does not support
lower atom-trimer channels and it is obviously not appropriate in
the dimer-dimer channel since the 3BDAA model treats the two 
dimers asymmetrically.
However, these channels are decisive for the tetramer width
since in their absence $\Gamma_{n,0}^d$ vanishes.
Thus, the failure of the 3BDAA model for $\Gamma_{n,0}^d$ 
is not surprising. Furthermore, 
the pointlike dimer approximation may only be reasonable when its size, 
being of order of $a$, becomes negligible as compared to $|A_d|$ 
(that determines also the size of the shallowest trimer if $A_d>0$), i.e.,
for $|A_d| >> a$. Under this condition the  3BDAA model
may provide a reasonable approximation of the full four-boson model
for particular observables like
 the atom-trimer elastic scattering cross section and 
the tetramer position $(b_n-B_{n,0}^d)$, but  not for 
reactions involving the dimer-dimer or  lower atom-trimer channels.
The intersection point estimations in Eqs.~\eqref{eq:ad} and \eqref{eq:at1}
refer to $|A_d|/a > 10^6$ and therefore should be reliable
whereas the 3BDAA model fails for $a_{n,0}^t/a_n^d$
as can be seen in in Fig.~\ref{fig:BG}.

\begin{figure}[!]
\includegraphics[scale=0.64]{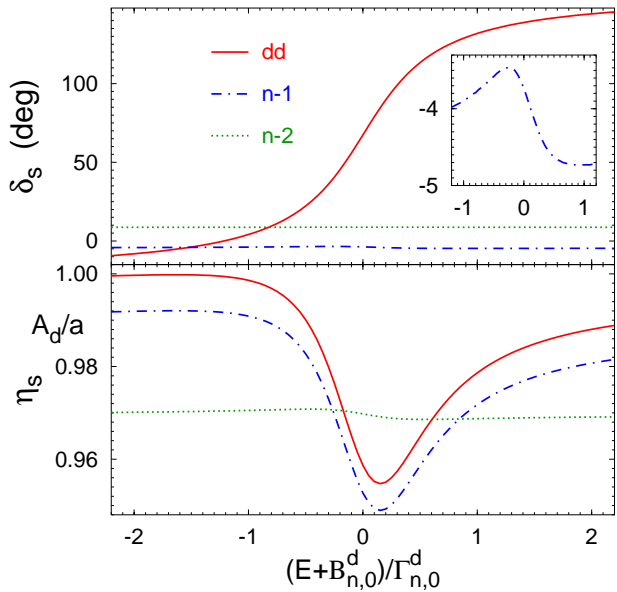} \quad\quad\quad
\caption{\label{fig:dd} (Color online)
$S$-wave phase shift and inelasticity parameter for the dimer-dimer 
(solid curves) and atom-trimer scattering where the trimer is
in the $(n-1)$th (dashed-dotted curves) or
in the $(n-2)$th (dotted curves) Efimov state.
The energy regime around the $(n,0)$th tetramer state
at $a_n^d/a = 0.8$ is shown. }
\end{figure}

The tetramer lies very close to the  atom-trimer threshold, even the
largest deviation around  $a_n^d/a = 0.8$ is only 
$(b_n - B_{n,0}^d)/b_n \approx 2.26 \times 10^{-4}$.
For comparison, the difference between the atom-trimer and 
dimer-atom-atom thresholds at this point is
$(b_n - b_d)/b_n \approx 8.86 \times 10^{-3}$.
Note that in a narrow interval around $a=a_{n,0}^t$
the tetramer is slightly above the atom-trimer threshold,  i.e., 
$B_{n,0}^d < b_n$. The width  $\Gamma_{n,0}^d$ vanishes at
$a=a_{n,0}^t$ while at $a > a_{n,0}^t$
the $(n,0)$th tetramer  becomes an inelastic virtual state (IVS)
 \cite{res_cpl}. Negative width $\Gamma_{n,0}^d < 0$
in the IVS case  implies the change of the energy sheet.
 The IVS corresponds to the pole of the transition operators $\mcu_{\beta\alpha}$
in the complex energy plane 
on one of the nonphysical sheets that is, unlike the one of UBS,
more distant from the physical sheet \cite{res_cpl}. The
IVS has a visible impact on the physical collision observables
only when it is located very near to the scattering threshold 
\cite{res_cpl}; an example referring to the four-boson system
can be found in Ref.~\cite{deltuva:11a}.
In contrast, the UBS always leads to a resonant behavior.
As example we show in Fig.~\ref{fig:dd}
the $S$-wave phase shift $\delta_S$ and inelasticity parameter  $\eta_S$
for the dimer-dimer and atom-trimer  scattering
at the total four-boson energy $E \approx -B_{n,0}^d$ and $a_n^d/a = 0.8$.
It is interesting to note that only the  dimer-dimer phase shift
increases by 180 deg while the inelasticity parameter
exhibits a rapid variation also for the collisions of atoms
and trimers in the $(n-1)$th Efimov state.
Thus, in these two cases the
inelastic reactions are significantly enhanced by the tetramer UBS 
whereas a pronounced resonant peak in the elastic cross section 
is present in the dimer-dimer channel only.
The $(n,0)$th tetramer has very little impact 
for the collisions of atoms
and trimers in the $(n-2)$th (and lower) Efimov state.

\begin{figure}[!]
\includegraphics[scale=0.64]{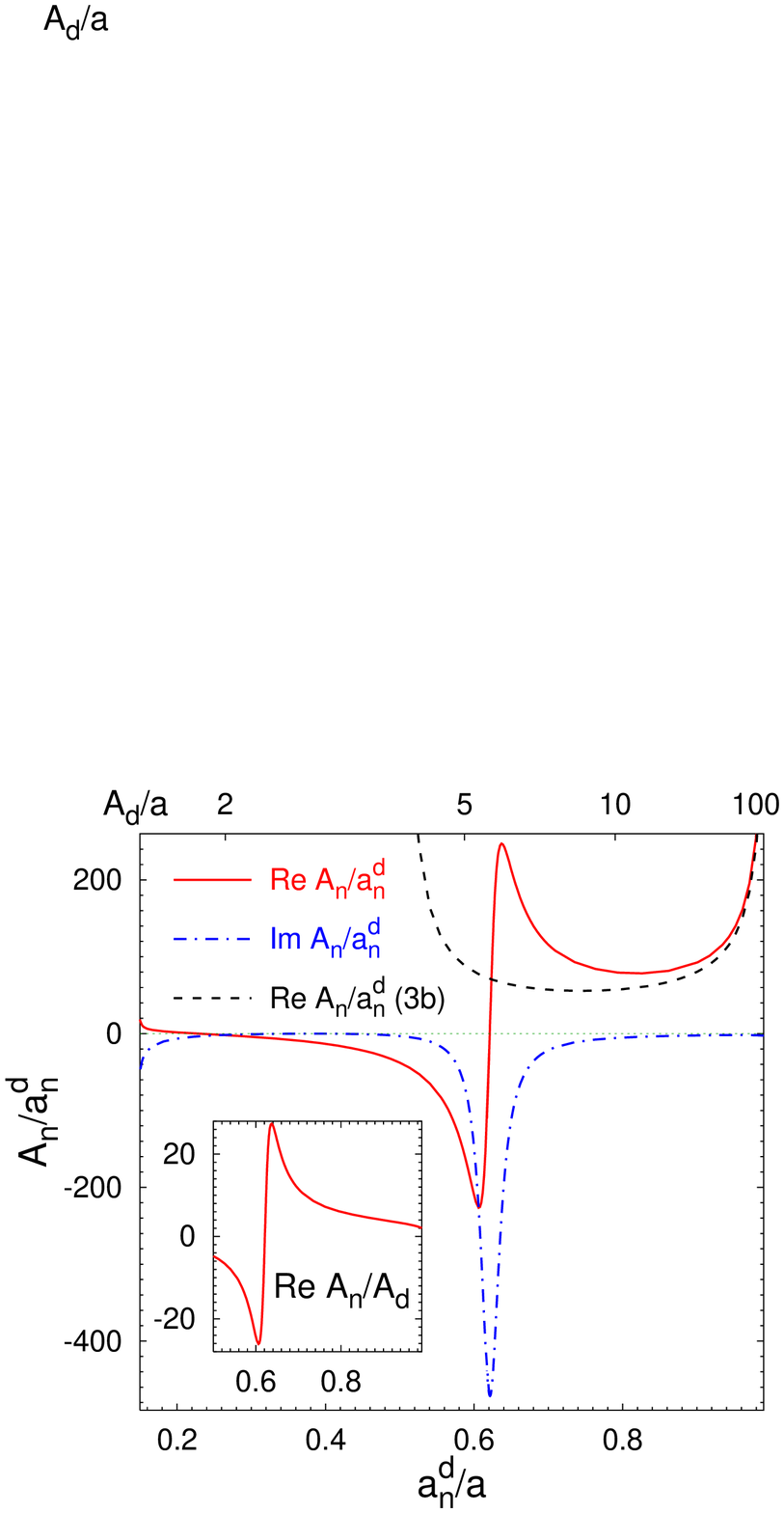}
\caption{  (Color online)
Real (solid curve) and imaginary (dashed-dotted curve) parts 
of the atom-trimer scattering length $A_n$
as functions of the two-boson scattering length $a$ and
the atom-dimer scattering length $A_d$. Results for
$\mathrm{Re}\,A_n$ obtained using the 3BDAA model are shown as dashed curve.
The thin dotted line represents the zero value. }
\label{fig:A}
\end{figure}

The $(n,0)$th tetramer intersection with the  $n$th atom-trimer threshold
manifests itself most prominently in the 
atom collisions with the $n$th  trimer at 
vanishing relative kinetic energy.
The corresponding atom-trimer scattering length $A_n$  is
shown in Fig.~\ref{fig:A} as a function of the 
two-boson scattering length $a$.
$A_n$ exhibits a resonant behavior 
with $|\mathrm{Im}\,A_n|$ having a peak at $a=a_{n,0}^t$.
The increase of $\mathrm{Re}\,A_n$ shown in Fig.~\ref{fig:A}
near  $a_{n}^d/a = 0.99$ is not related to the tetramers; it is due to
the scaling of the elastic cross section
with the spatial size of the trimer that grows
with increasing $A_d$  and decreasing $(b_n - b_d)$, i.e.,
$\mathrm{Re}\,A_n \sim A_d \sim (b_n - b_d)^{-1/2}$.
This is evident in the inset of Fig.~\ref{fig:A}
where the ratio $\mathrm{Re}\,A_n/A_d$ exhibits no 
resonant behavior near  $a_{n}^d/a = 0.99$. 
The inelastic reactions quantified by $\mathrm{Im}\,A_n$
are not significantly enhanced by the increased size of the trimer.
The failure of the 3BDAA model, especially for $\mathrm{Im}\,A_n$
and the resonant peak position, can be expected based on Fig.~\ref{fig:BG}
and its discussion. We show in Fig.~\ref{fig:A} the 3BDAA
prediction only for  $\mathrm{Re}\,A_n$ 
at larger $A_d/a$ values. Only at $A_d/a > 50$, i.e.,
$a_{n}^d/a > 0.96$, it agrees with the four-boson result
within 7\% or better. In the same regime the  3BDAA model underpredicts
 $\mathrm{Im}\,A_n$ almost by a factor of 40 as compared to the four-boson 
result. As estimated in Eq.~\eqref{eq:at1}
using the 3BDAA model, the $A_n$ resonance due to the 
$(n,1)$th tetramer is expected to take place much closer to $a=a_{n}^d$.
Finally we note that  rapid variations of $A_n$ near
 $a_{n}^d/a = 0.15$  are caused by the proximity of
the dimer-dimer threshold and the  Efimov tetramer
(of a different type)
that intersect the  $n$th atom-trimer threshold at
$a_{n}^d/a = 0.14730(1)$ \cite{deltuva:11b} 
and $a_{n}^d/a = 0.14706(1)$ \cite{deltuva:11a},
respectively. The $A_n$ evolution at
  $a_{n}^d/a < 0.15$ is presented in Ref.~\cite{deltuva:11a}.

\begin{figure}[t]
\includegraphics[scale=0.64]{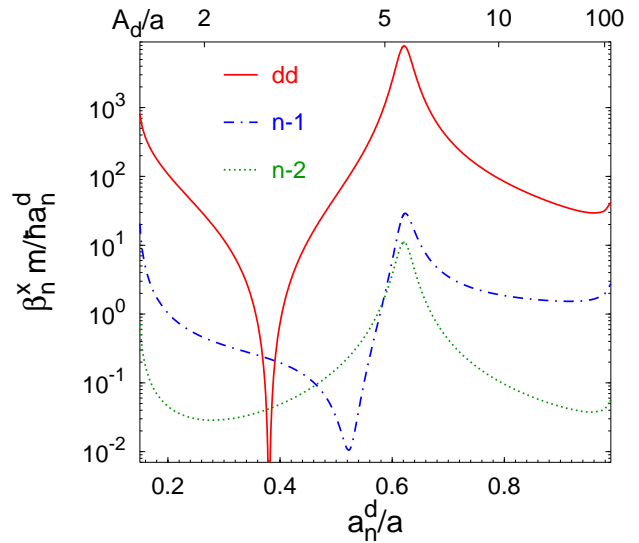}
\caption{ (Color online)
Inelastic reaction rates for vanishing relative energy
atom scattering from the $n$th trimer as functions of the two-boson 
scattering length $a$ or the atom-dimer scattering length $A_d$.
Rates for the dimer production (solid curve) 
and the trimer relaxation to the 
 $(n-1)$th (dashed-dotted curve) and to 
the $(n-2)$th (dotted curve) Efimov state are compared.}
\label{fig:rlx}
\end{figure}

The inelastic reaction rate for the atom scattering from the $n$th 
trimer in the ultracold limit is given \cite{deltuva:10c} by
$\beta_n = -(16\pi \hbar/3m) \mathrm{Im}\,A_n$ where $m$ is the
boson mass. 
 $\beta_n$ has contributions corresponding to the dimer production
$\beta_n^{dd}$ and the trimer relaxation $\beta_n^{n'}$ to more deeply
bound states $n'$, i.e.,
\begin{equation} \label{eq:rlx}
\beta_n = \beta_n^{dd} + \sum_{n'=0}^{n-1} \beta_n^{n'}.
\end{equation}
The respective dimensionless quantities $\beta_n^{dd} m / \hbar a_n^d$
and  $\beta_n^{n'} m / \hbar a_n^d$
are compared in Fig.~\ref{fig:rlx}. Consistently with our previous
findings for two-cluster collisions \cite{deltuva:10c,deltuva:11b}
and four-atom recombination  \cite{deltuva:12a},
in most of the considered
regime the dominating inelastic reaction is the one leading to the
most weakly bound final channel, i.e., the dimer-dimer channel
in the present case.
However, in a narrow interval around   $a_{n}^d/a = 0.38$  
an interesting and unusual situation takes place:
the dimer production rate $\beta_n^{dd}$ is strongly suppressed and
 the trimer relaxation process becomes the most important
inelastic reaction. Another exceptional situation can be seen
 around   $a_{n}^d/a = 0.52$  where $\beta_n^{n-1}$ has a minimum
such that $\beta_n^{n-1} < \beta_n^{n-2}$.

The considered regime is not yet explored experimentally. 
According to our predictions, the   $(n,0)$th tetramer
most clearly could be seen as a dimer production rate peak 
in ultracold collisions of atoms and excited trimers in the $n$th state,
since  $a_{n,0}^t$ is quite well separated from other special
values of $a$. Of course, in real experiments the  resonance position may
deviate from the universal value (\ref{eq:at}) 
due to finite-range effects; however, even without the dimer-trimer 
intersection, i.e., without the  $a_{n}^d$ point in Fig.~\ref{fig:ae},
the tetramer may still exist as it happens in our model with $n=1$.  
On the other hand, the observation of the resonant behavior in 
elastic or inelastic dimer-dimer collisions requires a very fine
energy resolution since the tetramer lies very close to the
atom-trimer threshold. 
The situation  is even more complicated
if  deeply bound dimer states are present such that the Efimov trimers
become UBS with finite width and lifetime.
According to Ref.~\cite{braaten:rev},
the trimer width is of the order $\eta_{*}b_n$ with $\eta_{*}$
being the atom-dimer inelasticity parameter;
for example, $\eta_{*} \approx 0.03$  in the experiment of 
Ref.~\cite{zaccanti:09a} performed at $a>0$ and 
 $\eta_{*} \approx 0.1$ in the experiment of
Ref.~\cite{ferlaino:09a} performed at $a<0$.
Thus, the width of the trimer greatly exceeds the distance $(B_{n,0}^d-b_n)$
between the tetramer and the atom-trimer threshold,
such that from the experimental point of view
the $n$th trimer and the $(n,0)$th tetramer are on top of each other.
Furthermore, in the presence of deep dimers the atom-dimer scattering length
 $A_d$ is complex with finite real and imaginary parts.
For example, $|\mathrm{Re}\,A_d/a| < 40$ at $\eta_{*} \approx 0.03$ 
 \cite{braaten:rev}. This should be sufficient for the existence
of the $(n,0)$th tetramer at $a$ slightly above $a_n^d$
but excludes all the others $(n,m\ge 1)$. For larger $\eta_{*}$
values even the  $(n,0)$th tetramer may be absent.
Thus, the experimental observation of the considered tetramers
would be extremely difficult.

\section{Summary \label{sec:sum}}

We studied bosonic tetramers that have approximate
dimer-atom-atom structure. They are unstable bound states in 
the continuum affecting the scattering processes in the four-boson system.
Exact four-particle equations  for the transition operators
were solved in the momentum-space  framework to
describe the four-boson collisions.
We accurately achieved the universal limit by considering reactions
involving highly excited trimers but, for a given trimer, the
results are restricted to the deepest tetramer only. 
We calculated the atom-trimer and dimer-dimer scattering observables
to extract the tetramer position and width. 
In particular, we determined the tetramer intersection with the 
atom-trimer threshold
and demonstrated that in  ultracold atom-trimer collisions 
it leads to a resonant enhancement of elastic and inelastic reactions.
We studied also the reliability of the three-body
dimer-atom-atom model for the atom-trimer collisions:
it may be reasonable for the elastic scattering near threshold 
and the real part of the tetramer energy at $|A_d| >> a$
but fails for the inelastic reactions and the tetramer width.



\begin{thebibliography}{10}

\bibitem{efimov:plb}
V. Efimov, Phys. Lett. B {\bf 33},  563  (1970).

\bibitem{kraemer:06a}
T. Kraemer~{\it et al}, Nature {\bf 440},  315  (2006).

\bibitem{braaten:rev}
E. Braaten and H.-W. Hammer, Phys. Rep. {\bf 428},  259  (2006).

\bibitem{lazauskas:he}
R. Lazauskas and J. Carbonell, Phys. Rev. A {\bf 73},  062717  (2006).

\bibitem{hammer:07a}
H.~W. Hammer and L. Platter, Eur. Phys. J. A {\bf 32},  113  (2007).

\bibitem{stecher:09a}
J. von Stecher, J.~P. D'Incao, and C.~H. Greene, Nature Phys. {\bf 5},  417
  (2009).

\bibitem{schmidt:10a}
R. Schmidt and S. Moroz, Phys. Rev. A {\bf 81},  052709  (2010).

\bibitem{yamashita:10a}
M.~T. Yamashita, D.~V. Fedorov, and A.~S. Jensen, Phys. Rev. A {\bf 81},
  063607  (2010).

\bibitem{deltuva:10c}
A. Deltuva, Phys.~Rev.~A {\bf 82},  040701(R)  (2010).

\bibitem{stecher:11a}
J. von Stecher, Phys. Rev. Lett. {\bf 107},  200402  (2011).

\bibitem{amado:73a}
R.~D. Amado and F.~C. Greenwood, Phys. Rev. D {\bf 7},  2517  (1973).

\bibitem{wang:12a}
Y. Wang, W.~B. Laing, J. von Stecher, and B.~D. Esry, Phys. Rev. Lett. {\bf
  108},  073201  (2012).

\bibitem{res_cpl}
A.~M. Badalyan, L.~P. Kok, M.~I. Polikarpov, and Y.~A. Simonov, Phys. Rep. {\bf
  82},  31  (1982).

\bibitem{deltuva:11a}
A. Deltuva, EPL {\bf 95},  43002  (2011).

\bibitem{dincao:09a}
J.~P. D'Incao, J. von Stecher, and C.~H. Greene, Phys. Rev. Lett. {\bf 103},
  033004  (2009).

\bibitem{deltuva:11b}
A. Deltuva, Phys.~Rev.~A {\bf 84},  022703  (2011).

\bibitem{deltuva:12a}
A. Deltuva, Phys.~Rev.~A {\bf 85},  012708  (2012).

\bibitem{ferlaino:08a}
F. Ferlaino, S. Knoop, M. Mark, M. Berninger, H. Sch\"obel, H.-C. N\"agerl, and
  R. Grimm, Phys. Rev. Lett. {\bf 101},  023201  (2008).

\bibitem{ferlaino:09a}
F. Ferlaino, S. Knoop, M. Berninger, W. Harm, J.~P. D'Incao, H.-C. N\"agerl,
  and R. Grimm, Phys. Rev. Lett. {\bf 102},  140401  (2009).

\bibitem{pollack:09a}
S.~E. Pollack, D. Dries, and R.~G. Hulet, Science {\bf 326},  1683  (2009).

\bibitem{zaccanti:09a}
M. Zaccanti, B. Deissler, C. D’Errico, M. Fattori, M. Jona-Lasinio, S.
  Müller, G. Roati, M. Inguscio, and G. Modugno, Nature Phys. {\bf 5},  586
  (2009).

\bibitem{grassberger:67}
P. Grassberger and W. Sandhas, Nucl. Phys. {\bf B2},  181  (1967); E. O. Alt,
  P. Grassberger, and W. Sandhas, JINR report No. E4-6688 (1972).

\bibitem{deltuva:07a}
A. Deltuva and A.~C. Fonseca, Phys.~Rev.~C {\bf 75},  014005  (2007).

\bibitem{deltuva:ef}
A. Deltuva, R. Lazauskas, and L. Platter, Few-Body Syst. {\bf 51},  235
  (2011).

\bibitem{alt:67a}
E.~O. Alt, P. Grassberger, and W. Sandhas, Nucl.~Phys. {\bf B2},  167  (1967).

\end{thebibliography}

\end{document}